\documentclass[sigconf]{acmart}
\usepackage{listings}
\usepackage{xcolor}
\usepackage{caption}
\usepackage{subcaption}
\usepackage{multirow}
\usepackage{lmodern}

\DeclareMathOperator{\td}{\textit{topicDist}}
\DeclareMathOperator{\ta}{\textit{topicAssign}}

\setcopyright{acmcopyright}
\copyrightyear{2025}
\acmYear{2025}
\acmDOI{XXXXXXX.XXXXXXX}

\acmConference[EASE 2025]{The 29th International Conference on Evaluation and Assessment in Software Engineering}{17–20 June, 2025}{Istanbul, Türkiye}

\begin{document}

\title{Towards Leveraging Large Language Model Summaries\\for Topic Modeling in Source Code}

\author{Michele Carissimi}
\email{m.carissimi4@campus.unimib.it}
\affiliation{%
\institution{Dept. of Informatics, Systems and Communication,\\
  University of Milano-Bicocca
  }
  \city{Milan}
  \country{Italy}
}

\author{Martina Saletta}
\email{martina.saletta@unibg.it}
\affiliation{%
  \institution{Dept. of Human and Social Sciences,\\
  University of Bergamo
  }
  \city{Bergamo}
  \country{Italy}
}

\author{Claudio Ferretti}
\email{claudio.ferretti@unimib.it}
\affiliation{%
  \institution{Dept. of Informatics, Systems and Communication,\\
  University of Milano-Bicocca
  }
  \city{Milan}
  \country{Italy}
}

\begin{abstract}
Understanding source code is a topic of great interest in the software engineering community, since it can help programmers in various tasks such as software maintenance and reuse. Recent advances in large language models (LLMs) have demonstrated remarkable program comprehension capabilities, while transformer-based topic modeling techniques offer effective ways to extract semantic information from text. This paper proposes and explores a novel approach that combines these strengths to automatically identify meaningful topics in a corpus of Python programs. Our method consists in applying topic modeling on the descriptions obtained by asking an LLM to summarize the code. To assess the internal consistency of the extracted topics, we compare them against topics inferred from function names alone, and those derived from existing docstrings. Experimental results suggest that leveraging LLM-generated summaries provides interpretable and semantically rich representation of code structure. The promising results suggest that our approach can be fruitfully applied in various software engineering tasks such as automatic documentation and tagging, code search, software reorganization and knowledge discovery in large repositories. 
\end{abstract}



\maketitle

\section{Introduction}

Understanding source code is a fundamental challenge in software engineering, since it helps programmers and, more in general, people that work on source code in various tasks such as code maintenance and code reuse. 

At the same time, topic modeling has proven to be an effective technique for extracting structured information from large text corpora. By identifying latent themes within a collection of documents, topic modeling enables applications such as document classification, content summarization, and knowledge discovery~\cite{churchill2022evolution,vayansky2020review}.

In the context of source code, topic modeling can help uncover high-level concepts, improve code organization, and help in mining large software repositories. Traditional topic modeling approaches, however, primarily rely on natural elements of code, such as identifiers and comments, which programmers introduce to enhance readability~\cite{biggerstaff1993concept,kuhn2007semantic,mahmoud2017semantic}. This dependence can limit their effectiveness when such elements are missing, inconsistent, or poorly maintained.

In this paper, we propose a novel approach that combines the ability of LLMs to comprehend source code and perform document embeddings, together with topic modeling techniques to extract meaningful topics from code. Specifically, we choose Python code and we begin by processing functions where comments and docstrings are removed, and function names are obfuscated to eliminate explicit semantic clues. We then use an LLM to generate summaries of these functions, capturing their key functionalities. Finally, we apply topic modeling to this corpus of LLM-generated summaries, allowing us to infer meaningful topics even in the absence of traditional natural-language cues. This approach offers a robust method for analyzing and organizing code, with potential applications in program comprehension tasks such as automatic documentation, software reorganization, code search, and knowledge discovery in large repositories.

Data and code for replicating our experiments are publicly available\footnote{\url{https://zenodo.org/records/15036066}}.

\section{Summary-based Topic Modeling}
\label{sec:overview}

Recent advances in large language models (LLMs) have demonstrated remarkable capabilities in several domains, and their potential is also evident in the program comprehension field~\cite{nam:llmcode}. Besides, transformer-based topic modeling techniques have proven effective in extracting semantic information from text~\cite{reuter2025probabilistic}. 

This paper explores a novel approach that integrates these findings to automatically identify meaningful topics in a corpus of Python programs.

In the literature, there exist studies that explored the application of topic modeling techniques to source code, but all these approaches primarily rely on the ``natural elements'' (e.g. identifiers, comments, and docstrings) that programmers introduce to enhance readability and maintainability~\cite{iammarino2020topic,kuhn2007semantic,biggerstaff1993concept,saeidi2015itmviz}.

\begin{figure*}[htb]
    \centering
    \includegraphics[width=
    .8\textwidth]{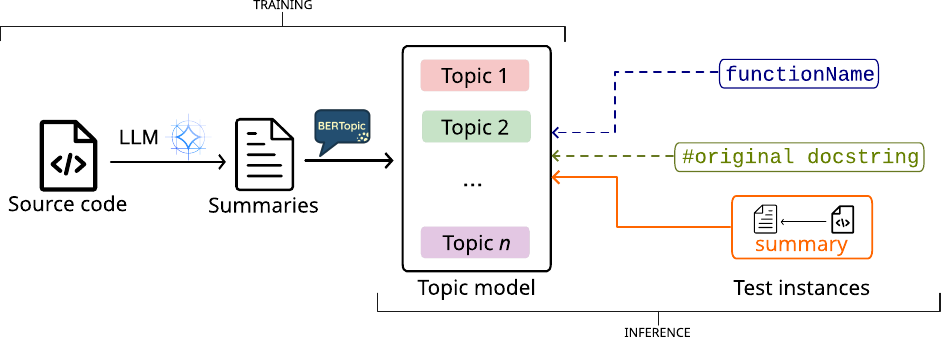}
    \caption{Study overview. Source code summaries generated by an LLM are used to identify the latent topics. This topic model can be used for inferring code information from different code representations.}
    \label{fig:overvoew}
\end{figure*}

In this study, we aim to define a topic modeling technique for source code that remains effective even when such natural elements are removed. To this end, we (1) start with Python functions where comments and docstrings are removed and function names are obfuscated to eliminate any explicit hints about their purpose. We then (2) leverage a large language model (LLM) to generate summaries of these functions, capturing their underlying semantics without relying on manually introduced textual cues. Finally, (3) we apply topic modeling to this corpus of LLM-generated summaries to extract meaningful topics, as outlined in Figure~\ref{fig:overvoew}. This approach allows us to show whether topic modeling can effectively uncover the structure and organization of code even in the absence of conventional natural-language annotations.

In a topic modeling context, for each document $d$ of the corpus $D$, the model $M$ defines a function $\td_M \colon D \to \mathbb{R}^{n}$ that returns the topic distribution for the given document in the form of a probability distribution, with $n$ representing the total number of modeled topics. In other words, for each $d \in D$, $\td_M(d) = (p_{d_1}, \dots, p_{d_{n}})$ and $p_{d_1} + p_{d_2} + \dots + p_{d_{n}} = 1$.
Similarly, each topic $T_i \in \{T_1,\dots,T_n\}$ is represented through a vector of weights $T_i = (w_{i,1}, w_{i,2}, \dots,w_{i,|W|})$, where each $w_{i,k}$ represents the relevance that the word $k$ has for the topic $T_i$ and $|W|$ is length of the vocabulary. Finally, we associate each document $d$ with a topic, where a topic is indexed with an integer $1 \leq i \leq n$, with the function $\ta_M \colon D \to \{1, \dots, n\}$, defined as $\ta(d) = i$ if $p_{d_i} = \max(p_{d_1}, \dots, p_{d_{n}})$.

To validate our approach, we trained two topic models $M_{\textit{summ}}$ and $M_{\textit{doc}}$ starting from different representations of the same instances: function summaries produced by an LLM for $M_{\textit{summ}}$ and docstrings for $M_{\textit{doc}}$, respectively. Notice that function summaries are obtained starting from functions in which comments have been removed and the name of the function obfuscated, so as to remove the intrinsic code naturalness~\cite{allamanis:bigcode}. 

According to this notation, we now introduce the four metrics we used to assess the robustness of our results. We consider the docstrings as the golden standard, and we measure how the inference on $M_{\textit{doc}}$ differ when made on the code summaries or on the original function names. In other words, we define four distances between the inferences of a model. 

\paragraph{$d_{MSE}$} This distance is based on the mean squared error between the probability distributions associated with the topic assignment to each document.
Given a model $M$ with $n$ topics, we define the distance between topics of two documents $\td_M(d) = (p_{d_1}, \dots, p_{d_{n}})$ and $\td_M(d') = (p_{d'_1}, \dots, p_{d'_{n}})$ as: 
\[
d_{MSE}(d,d') = \frac{1}{n} \sum_{i=1}^{n} \bigl(p_{d_i} - p_{d'_i}\bigr)^2
\] 
\paragraph{$d_{TOP}$} Given two topic distributions inferred for documents $d$ and $d'$, $d_{TOP}(d,d')$ is the  number of common topics between the 10 most probable of each topic distribution; a higher value is associated to a better similarity between the topic distributions,
\paragraph{$d_{TOPw}$} Given two topic distributions inferred for documents $d$ and $d'$, $d_{TOP}(d,d')$ is the average cosine-similarity between the 10 most probable topics of the topic distributions, as represented by the respective rows in the topic-term relevance matrix defined by the topic model; a higher value is associated to a better similarity between the topic distributions,
\paragraph{$d_\cap$} Given assigned topics $\ta(d)$ and $\ta(d')$, $d_{TOP}(d,d')$ is the number of common words between the 5 most relevant words of each topic; a higher value is associated to a better similarity between the topics; notably, this metric can be computed even between topics assigned by different topic models.

Our results are promising. From a qualitative perspective, this is evident in Table~\ref{tab:topics}, where we can observe meaningful and well-separated topics. To further assess the effectiveness of our approach, we compare, by using the above defined distance measures, the obtained topics with those derived from existing docstrings, showing that our method achieves similar scores across different evaluation metrics. Finally, we demonstrate that when using only function names, the results are significantly worse. This indicates that our approach can successfully identify and model topics from source code by relying solely on code structure rather than leveraging natural-language elements.

\section{Experimental Settings} 
The experiments described in this study have been carried out on a set 10.000 Python functions, randomly sampled from the Python partition of CodeSearchNet~\cite{CodeSearchNet}, a popular dataset widely used in the software engineering community that includes a large corpus of source code
snippets and their corresponding docstrings.

The records in the dataset are JSON objects, and for each instance we considered the following fields: 
\begin{itemize}
  \item \texttt{func\_name}: the attribute representing the name of the function; 
  \item \texttt{whole\_func\_string}: the attribute representing the source-code snippet; 
  \item \texttt{func\_documentation\_string}: the attribute representing the docstring for each function. 
\end{itemize}
As fully motivated in Section~\ref{sec:overview}, prior to asking the LLM to provide the summaries, all function names in their declarations were replaced with a fixed, randomly generated, placeholder string, to eliminate any semantic information that could reveal the purpose of the functions. This was achieved by using a regular expression to identify and replace the function names in their definition lines. This step ensured that the LLM focused solely on understanding the logic within the source code rather than leveraging potentially meaningful naming conventions.
This preprocessing step not only ensured that the LLM's summarization focused purely on the logic of the source code, but also enabled a later comparison between the topic distributions inferred from: 
\begin{enumerate}
\item summarizations generated by the Gemma~\cite{gemma} model from the processed source code
\item tokenized function names
\item original docstrings
\end{enumerate}
%
Additionally, with the same goal, all the comments were removed from the source code snippets during preprocessing with the use of regular expressions to replace comments with empty strings. This was done to eliminate auxiliary information that might simplify the summarization task, ensuring that the LLM's output relied purely on the syntactic and structural elements of the code.

In order to query the LLM, we constructed the prompt to be used by concatenating a fixed portion, which explains the task and how to format the output, and a variable portion, containing the source-code snippet to be explained.
\begin{itemize}
    \item \textbf{Base query:} \texttt{``Consider the following source code and provide a description of its purpose.''}
    \item \textbf{Prompt template:} \texttt{base\_query + source\_code + `` The output should follow this format: \#\#\#\#\# Description: <source code description>''}
\end{itemize}
The source code description generation process was conducted using the Gemma2 2B-it model~\cite{gemma,Gemma2-2B}, a 2-billion-parameter language model designed for text generation. The maximum token limit for generated descriptions was set to 1024 to generate concise and comprehensive outputs. Queries were executed via Huggingface's pipeline API \footnote{https://huggingface.co/docs/transformers, last visit November 2024}, which provided an interface for integrating the model into the experimental pipeline. The choice of Gemma2 2B-it was driven by its balance of computational efficiency and performance.

The resulting data underwent two preprocessing steps. First, all text was converted to lowercase, and numbers, punctuation marks, and commonly defined stopwords were removed. The stopwords were identified using the default collection provided by the NLTK library \cite{NLTK}, which is widely used for natural language processing tasks.

Next, the records were analyzed to identify words appearing in at least 75\% of documents. These high-frequency words were removed under the assumption that they carry limited semantic value and contribute minimally to distinguishing topics. By excluding these terms, the preprocessing ensured that the corpus retained only those words more likely to provide meaningful insights during the topic modeling process.

BERTopic~\cite{BERTopic} is a topic modeling algorithm that uses embeddings from transformer-based models combined with clustering techniques to group semantically related documents. Unlike traditional methods like Latent Dirichlet Allocation (LDA)~\cite{blei:lda}, which rely on term frequency distributions, BERTopic utilizes vector representations of text to capture contextual relationships between words and phrases. These embeddings are reduced in dimensionality using algorithms such as UMAP~\cite{mcinnes:umap}, enabling efficient clustering and topic extraction. This approach is particularly useful for analyzing large text corpora where semantic meaning is essential for identifying coherent topics.

After preprocessing, the dataset was used as a corpus for the topic modeling process. BERTopic was applied to identify coherent topics within the dataset. The dimensionality reduction step, a crucial part of BERTopic's pipeline, was performed using UMAP. All the experiments are performed with the following parameters:

\begin{itemize}    
    \item \texttt{nr\_topics = 40} and \texttt{topic\_size = 25} to avoid topics having few associated documents 
    \item \texttt{metric = 'cosine'} to measure the distance among documents in the embedding space
    \item \texttt{nr\_neighbors = 10} and \texttt{min\_distance = 0.01} for the dimensionality reduction with UMAP
\end{itemize}

Selecting the optimal number of topics in topic modeling is an open problem, as there is no universally accepted method for determining this parameter. For this study, the number of topics (\texttt{nr\_topics}) was set to 40 after running the model multiple times with varying values and evaluating the results. The decision was guided by a balance between coherence score, measuring the internal consistency of the topics, and interpretability, ensuring that the topics identified meaningful patterns within the corpus. This iterative approach allowed the selection of a number that provided both statistically robust and human-interpretable results.

The initial dataset, composed of 10.000 records, was split creating a 9.500 records train set and a 500 records evaluation set. 
After applying the topic modeling process, BERTopic generated a series of topics. Each topic is represented by a ranked list of words from the corpus, sorted by their relevance to that topic. This ranking provides insight into the defining terms for each topic, aiding in interpretation.

To associate topics to a document, in what is considered the inference step, BERTopic computes a probability distribution across all modeled topics, indicating the likelihood of each topic being relevant to the document. Also, a specific topic is assigned if its probability is sufficiently high relative to the others. If no topic meets this threshold, the document is labeled as "-1" or "outlier," indicating that it does not fit well into any single topic. This mechanism ensures that only documents with strong associations to a specific topic are classified, while ambiguous or unrelated documents remain unassigned.

\section{Results}
\label{sec:Results}
The results of the topic modeling process were evaluated using multiple metrics, with a primary focus on the coherence score, which quantifies the semantic consistency among the most relevant words within each topic. A low coherence score indicates that the top words are semantically distant, making the topic harder to interpret. In contrast, a high coherence score signifies strong semantic connections among the top words, leading to topics that are easier to understand and label.
In this study, the $C_v$ coherence score was utilized. This metric combines a sliding window approach, word co-occurrence counts, and vector representations of words to measure cosine similarity among the top words in a topic~\cite{roder:coherence,syed:coherence}. By leveraging both statistical co-occurrence and semantic similarity, $C_v$ provides a robust evaluation of topic interpretability, making it suitable for assessing the quality of topics derived from domain-specific datasets.

A list of topics resulting from our modeling process are reported in \autoref{tab:topics}. The table shows the five most relevant words for each topic, along with a tentative title that summarizes the topic scope and the coherence score of each topic. 
In the table, Topic 32 has a coherence value far lower than all others, but even there the top 5 words representing it all belong to the same domain related to code managing colors.

\begin{table*}[htb]
\small
\caption{Topic models}
\label{tab:topics}
\centering
\begin{tabular}{|l|l|l|c|}
\hline
\textbf{Topic} & \textbf{Top words} & \textbf{Label} & \textbf{Coherence}\\
\hline
0 & request, url, respnse, api, http & HTTP Communication & 0.89 \\
\hline
1 & message, device, network, system, address & Network Communication & 0.58 \\
\hline
2 & array, calculates, matrix, calculation, values & Matrix Operations & 0.73\\
\hline
3 & myclass, attribute, instance, named, within & Object-oriented Programming & 0.53\\
\hline
4 & path, directory, files, module, filename & Filesystem & 0.69\\
\hline
5 & database, query, import, table, sql & Database & 0.45 \\
\hline
6 & configuration, yaml, commandline, yang, arguments & Configuration Files & 0.39\\
\hline
7 & csv, gene, bioseqfeature, featuretype, sequence & Genomic Data & 0.41 \\
\hline
8 & time, date, datetime, dateutilparser, parse & Date and Time Parsing & 0.70\\
\hline
9 & image, pil, images, pixel, color & Image Processing & 0.52 \\
\hline
10 & model, layer, training, learning, tensor & Neural Networks & 0.88 \\
\hline
11 & log, logging, message, exception, logger & Logging & 0.48\\
\hline
12 & widget, window, gui, layout, button & GUI Components & 0.53\\
\hline
13 & dictionary, key, keys, keyvalue, dictionaries & Dictionaries (data structures) & 0.64\\
\hline
14 & pattern, regular, string, match, word & Regular Expressions & 0.79\\
\hline
15 & plot, axes, import, plots, astropyvisualization & Data Visualization & 0.64\\
\hline
16 & byte, bytes, encoding, string, encoded & Data Encoding & 0.74\\
\hline
17 & node, tree, child, children, parent & Trees (data structures) & 0.85\\
\hline
18 & typing, import, dict, future, optional & Type Annotations & 0.51\\
\hline
19 & description, string, information, format, field & Field Descriptions & 0.47\\
\hline
20 & hash, key, signature, algorithm, password & Cryptographic Hashing & 0.81\\
\hline
21 & graph, nodes, node, edges, edge & Graphs (data structure) & 0.84\\
\hline
22 & package, repository, dependencies, packages, git & Package Management & 0.64\\
\hline
23 & xml, element, callback, config, configuration & XML Configuration & 0.60\\
\hline
24 & html, template, content, element, rended & Websites & 0.49\\
\hline
25 & decorator, decorated, original, wrapper, acts & Function Decorators & 0.64\\
\hline
26 & json, dictionary, python, string, path & JSON & 0.49\\
\hline
27 & command, commands, output, shell, execute & Shell Commands & 0.84\\
\hline
28 & dataframe, columns, column, pandas, rows & Pandas Dataframes & 0.69\\
\hline
29 & aws, region, bucket, service, profile & AWS (Amazon Web Services) & 0.65\\
\hline
30 & table, row, column, cell, rows & Tables & 0.75\\
\hline
31 & cache, caching, cached, torrent, key & Caching & 0.47\\
\hline
32 & color, rgb, colormathcolorconversions, colormathcolorobjects, hscolor & Colors & 0.08\\
\hline
33 & series, time, percentile, calculates, index & Time Series & 0.54\\
\hline
34 & filter, filters, filtering, criteria, filtered & Filters & 0.50\\
\hline
35 & version, string, prerelease, versions, major & Versioning & 0.42\\
\hline
36 & language, translation, translations, languages, header & Language Translation & 0.37\\
\hline
37 & email, user, roles, address, permissions & Email Management & 0.38\\
\hline
38 & download, downloaded, downloads, directory, url & File Downloads & 0.75\\
\hline

\end{tabular}
\end{table*}

\begin{figure}[htb]
    \centering
    \includegraphics[width=\linewidth]{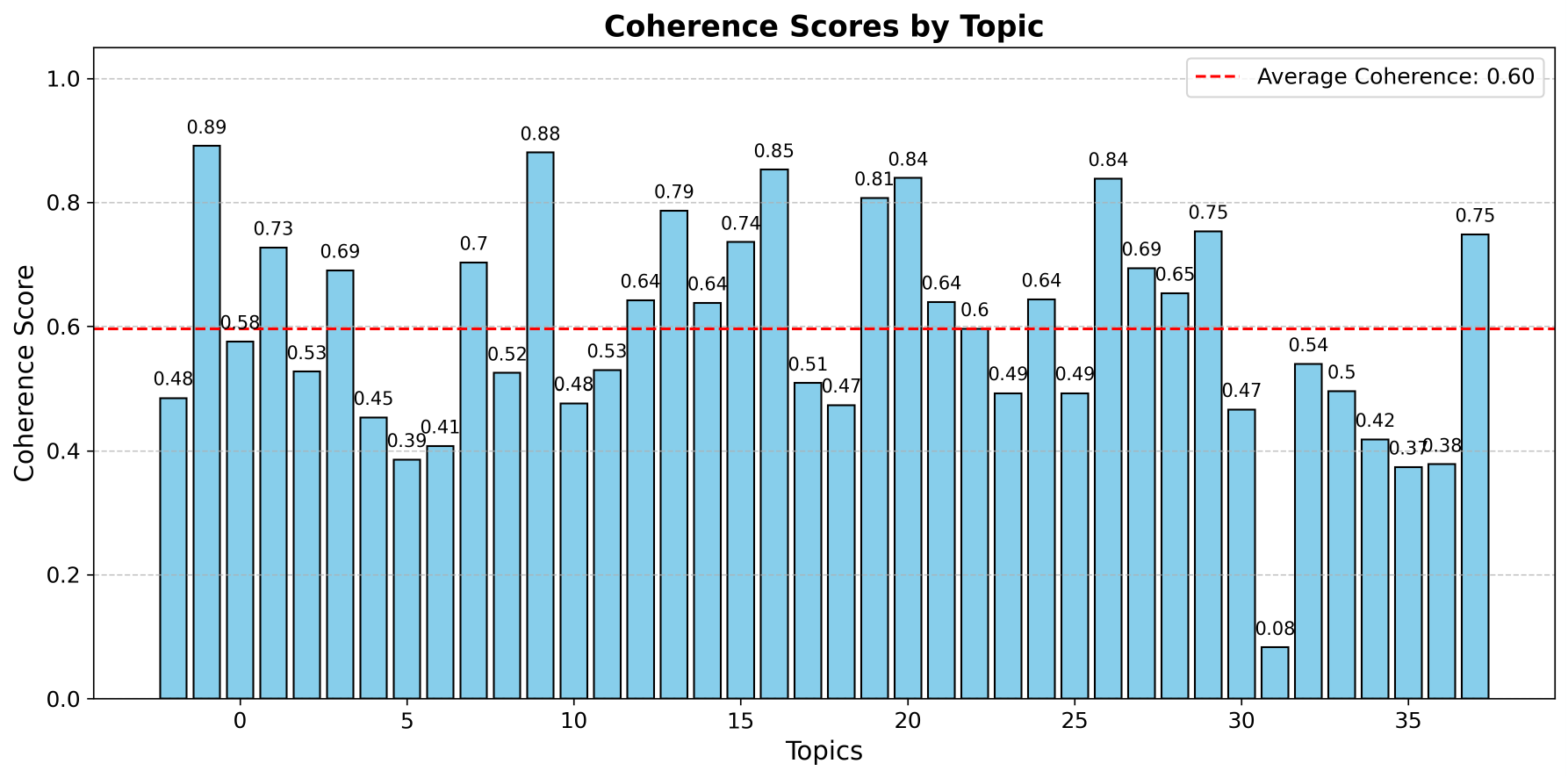}
    \caption{Coherence scores}
    \label{fig:CoherenceScores}
\end{figure}

The coherence scores for all topics are also visualized in~\autoref{fig:CoherenceScores}, which also highlights the average coherence score across topics. The average coherence score was approximately 0.60, indicating moderate interpretability and the presence of meaningful word clusters.
The coherence score of 0.60 suggests that the topics identified by the model are both relevant and interpretable. This level of coherence reflects the model's ability to extract meaningful patterns from a highly technical and structured corpus.
We also computed the average coherence score of the topics modeled from the set of original docstrings, which we can consider as a semantically rich content, but this value resulted to be only 0.38, thus suggesting the value of our result.  

\begin{figure*}[htb]
    \centering
    \includegraphics[width=.8\linewidth]{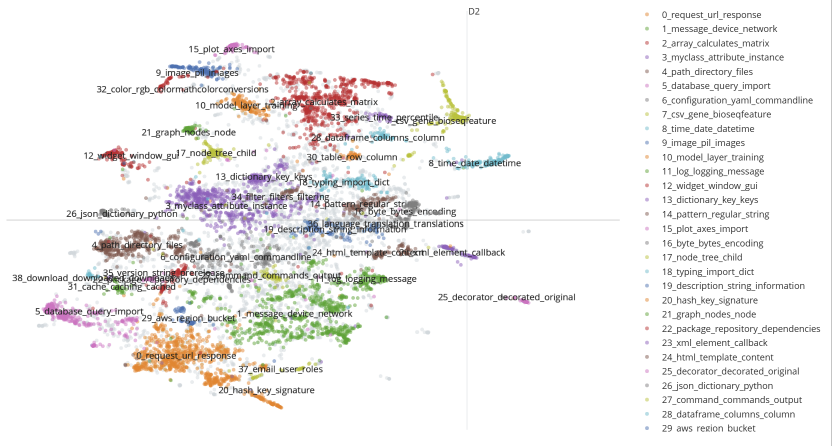}
    \caption{Document embeddings visualization. Each point represents a code instance, with colors indicating the assigned topics.}
    \label{fig:DocsVisual}
\end{figure*}

Documents and their associated topics can also be visualized in a two-dimensional space (\autoref{fig:DocsVisual}), where each point represents a document. The positioning of the points reflects their semantic embeddings, generated through dimensionality reduction using UMAP~\cite{mcinnes:umap}. Semantic similarity between documents is indicated by their proximity, while the color assigned to each document encodes the most relevant topic for that document.

The presence of dense and well-separated clusters indicates that the model effectively grouped semantically similar documents, suggesting that the identified topics are coherent and interpretable. Some overlap between clusters or the presence of scattered points reflect the nature of the data, where a single document can be associated to more than one strongly relevant topic. In fact, as detailed in Section~\ref{sec:overview}, the topic inference produces, for each document, a probability distribution over the whole set of modeled topics.

\begin{table}[htb]
\caption{Comparison between reference topics and topics inferred under two models and on different inputs (metrics from Section~\ref{sec:overview}, discussion in Section~\ref{sec:Discussion}).}
\label{tab:compare}
\centering
\begin{tabular}{l|l|llll}
\hline
& & $d_{MSE}$ & $d_{TOP}$ & $d_{TOPw}$ & $d_\cap$\\
\hline
\multirow{2}{1cm}{$M_{\textit{doc}}$} & Summaries & $0.013$ & \boldmath$3.46$ & \boldmath$0.54$ & \boldmath$3.71$\\
& Names & $0.013$ & $3.11$ & $0.51$ & $2.18$\\
\hline
$M_{\textit{summ}}$ & Summaries & N/A & N/A & N/A & $2.37$\\
\hline
\end{tabular}
\end{table}

\section{Discussion}
\label{sec:Discussion}

We assess the value of our approach according to different usage contexts.
The key element is the exploitation of the internal representation of an LLM in order to create semantically rich natural language descriptions of input source code.
Previous proposals regarding the creation of topic models for source code (\cite{biggerstaff1993concept,kuhn2007semantic}) required the availability of natural language elements in the source code, such as comments or meaningful identifier.
We move along a different way, since topics are modeled based on the natural text generated by the LLM from the input source code.
This opens to applications to source code without accompanying documentation, or without useful identifiers.
Consistently, we experimented our tool always on code where we removed any comment and obfuscated the functions' identifiers.

The method can be evaluated in terms of internal quality of the topics it creates, and we showed in Section~\ref{sec:Results} that we generate topics with high coherence measure.
But also, the method exhibits advantageous performance when compared against techniques based on natural language elements from source code.
To this aim, we can consider two main reference usage contexts:
one where only the code is available, and one with the availability of developer authored documentation.
In these two settings, the value of the topics generated or assigned with our technique has been assessed in terms of their alignment with the topics modeled based on the documentation and then assigned to some code from its own documentation. We consider these latter topics the reference target.
The results, presented in Table~\ref{tab:compare}, show that the summarization produced by the LLM are always close to the reference topics, and therefore it compensates the lack of documentation.

In detail, in Table~\ref{tab:compare} we show the average results produced according to the metrics defined in Section~\ref{sec:overview} when comparing topics inferred under different settings against the reference target topics.
The first two lines evaluate the topic inferences produced on the model $M_{\textit{doc}}$ built from docstrings, but first taking as input the summaries generated by the LLM and then taking as input the tokenized function names, respectively.
The numbers show that the topics based on LLM-generated summaries are more similar to reference topics
than the ones generated from the function names and their natural semantic.
Interestingly, a remark can be done about the apparent effectiveness of inferring topics starting from function identifiers.
It is good practice that the developer coding a function assigns to it a meaningful and representative identifier.
As suggested by~\cite{allamanis:bigcode}, the naturalness of those identifiers is a rich information,
based on which machine learning systems can be trained to successfully perform tasks such as
automatic function name suggestion.
On the other hand, it has been shown that when summarizing source code, the removal of natural content leads to a significant drop in performance~\cite{ferretti:icpc}.
This can explain the quality we measured for topics generated from identifiers when comparing them
to the reference topics generated from the documentation.
Nonetheless, the topic defined from the summary texts generated by the LLM
in our method appear to be better.
Finally, in the last row of Table~\ref{tab:compare} we evaluate against reference topics
the ones inferred from generated summaries, but in this case the topic model is $M_{\textit{summ}}$, which has been built from summaries itself.
In this setting only the $d_\cap$ metric could be applied, since the available topics are not
identical to the topics of the model built from docstrings, in terms of vocabulary and associated words relevance values.
The similarity to the reference topic assignment is still good, with an average number
of $2.37$ (over $5$) top relevant terms shared with the top terms of the reference topic.

Overall, our approach offers a performance which is close to that of topic modeling with code documentation available,
but more robust in settings where the code is lacking good comments or meaningful function identifiers.

\section{Conclusion and Further Directions}

This work introduces a method to build topic models on source code corpora without requiring the availability of natural text (comments or function identifiers), by leveraging the capability of LLMs to generate natural language summaries for code snippets. The new method has better performance, according to several metrics we chose, than more traditional techniques requiring comments and documentation to build topics.

Further research could be devoted to tailoring and applying this method to other source code knowledge extraction tasks.

\bibliographystyle{ACM-Reference-Format}
\bibliography{main}

\end{document}